\documentclass[final]{ws-ijmpd}

\usepackage[OT2,OT1]{fontenc}
\usepackage{amsfonts}
\usepackage{amssymb}
\usepackage{graphicx}
\usepackage{bm}

\def\be{\begin{equation}}
\def\ee{\end{equation}}
\def\ba{\begin{eqnarray}}
\def\ea{\end{eqnarray}}
\def\bs{\begin{subequations}}
\def\es{\end{subequations}}

\def\rme{e}
\def\rmd{d}
\def\rmi{i}
\def\p{\partial}

\def\B{\Box}
\def\a{\alpha}

\def\s{\sigma}

\def\g{\gamma}

\def\k{\kappa}

\newcommand{\Eq}[1]{(\ref{#1})}

\begin{document}


\title{COSMOLOGICAL ROLLING SOLUTIONS OF\\ NONLOCAL THEORIES}

\author{GIANLUCA CALCAGNI}
\address{Institute for Gravitation and the Cosmos, Department of Physics,\\ The Pennsylvania State University, 104 Davey Lab, University Park,\\ Pennsylvania 16802, USA}
\author{GIUSEPPE NARDELLI}
\address{Dipartimento di Matematica e Fisica, Universit\`a Cattolica,\\
via Musei 41, 25121 Brescia, Italia}
\address{INFN Gruppo Collegato di Trento, Universit\`a di Trento,\\ 38100 Povo (Trento), Italia}

\maketitle

\begin{abstract}
We find nonperturbative solutions of a nonlocal scalar field equation, with cubic or exponential potential on a cosmological background. The former case corresponds to the lowest level effective tachyon action of cubic string field theory. While the well known Minkowski solution is wildly oscillating, due to Hubble friction its cosmological counterpart describes smooth rolling towards the local minimum of the potential.
\end{abstract}
\date{April 20, 2009}

\rightline{IGC-09/4-3\hspace{8.2cm} arXiv:0904.4245}


\section{Introduction}

The search for signatures of nonperturbative quantum gravity in the history of the early universe has motivated several authors to consider cosmological models inspired by open string field theory (OSFT),\cite{are04,AJ,AKV1,cutac,AK,AV,kos07,AJV,AV2,cuta2,Jou07,Jo081,ArK,Jo082,NuM,KV} the $p$-adic string,\cite{Jo082,NuM,BBC,lid07,BC,cuta4,BK2} or other nonlocal effective actions.\cite{SW,BMS,kho06,DW,NO,Jhi08,CENO,DeW} In the first category, the typical starting point is a nonlocal scalar equation with polynomial potential,
\be\label{geom}
\left(\B+m^2\right)\tilde\Phi+\s \Phi^n=0\,,
\ee
where $\B$ is the d'Alembertian operator, $m$ and $\s$ are constants, and $\tilde\Phi\equiv \rme^{s\B}\Phi$. This type of pseudo-differential operators can be conveniently manipulated with the diffusion equation approach, which was developed and employed, in diverse numerical or analytical guises, in Refs.~\cite{cuta2,Jou07,Jo081,Jo082,NuM,cuta4,vol03,FGN,vla05,roll,cuta3,MuN,cuta5}. In brief, the method can be outlined as follows:\cite{cuta2,cuta4,roll,cuta3} (A) Let the scalar field $\phi(t)\to\phi(r,t)$ evolve also along an auxiliary direction $r$ and interpret $s$ as a fixed value of the extra variable. (B) Find a solution $\phi(0,t)$ of the \emph{local} system ($s=0$). This is the initial condition for a system that evolves in $r$ via the diffusion equation
\be\label{difeq}
(\B+\p_r)\Phi=0\,.
\ee
(C) The solution of the diffusion equation is
\be\label{solde}
\Phi(r,t)=\rme^{r\B}\Phi(0,t)\,.
\ee
In particular, the effect of the nonlocal operator $\rme^{q\B}$ is a shift of the auxiliary variable $r$:
\be\label{q}
\rme^{q\B}\Phi(r,t)= \rme^{-q\,\p_r} \Phi(r,t)=\Phi(r-q,t)\,.
\ee
(D) If the initial condition is chosen to be constant almost everywhere, the final configuration $\Phi(s,t)$ obtained by diffusion along $r$ is a smooth function which solves (exactly or approximately) the original nonlocal system for some $s$ (here we do not consider smooth initial conditions\cite{cuta2,roll}).

In a recent paper,\cite{cuta5} we found a rolling homogeneous solution to \Eq{geom}:
\be\label{phi}
\Phi=\Phi(\a,z)=\frac{\g\left(\a,z^2\right)}{\Gamma(\a)}\,,
\ee
where $z\equiv 2\sqrt{r}\,t$ and
\be
\gamma(\a,z^2)=2\int_0^z \rmd \tau\, \tau^{2\a-1} \rme^{-\tau^2}
\ee
is the lower incomplete gamma function (see Sec.~8.35 of Ref.~\cite{GR}). The constant $\a\geq 0$ will determine the shape of the scalar field profile. First, we derived an approximate scaling formula which expresses powers of the incomplete gamma function as a single $\g$ with rescaled arguments:
\be\label{dupli2}
[\Phi(\a,z)]^n \approx \Phi(n\a,\kappa z)\,,
\ee
where
\be
\k=\kappa(n,\a)\equiv \frac{\left[\Gamma(n\a+1)\right]^{1/{(2n\a)}}}{\left[\Gamma(\a+1)\right]^{1/{(2\a)}}}\,.\label{kappa2}
\ee
When $n=1+1/\a$, using the properties of $\g$ one can show that \Eq{phi} approximately (in the sense of \Eq{dupli2}) solves \Eq{geom} with
\be
s=\left(1-\frac{1}{\k^2}\right)r\,,\qquad m^2= -\frac{\a\k^2}{r}\,,\qquad \s=-m^2\,,
\ee
and the universe expands as a power-law:
\be
\B=-\p_t^2 -\frac{1-2\a}{t}\p_t\,.
\ee
The solution obeys the diffusion equation \Eq{difeq} and the action of the nonlocal operator in \Eq{geom} is a rescaling:
\be
e^{s\B}\Phi(\a,z)=\Phi(\a,\k z)\,.
\ee
The level of approximation of the solution \Eq{phi} is determined by the duplication formula \Eq{dupli2} and is sufficient to construct easily instantonic tachyon solutions of supersymmetric string field theory on flat Minkowski spacetime, $(n,\a)=(3,1/2)$. The approximation scheme related with the duplication formula was extensively discussed in Ref.~\cite{cuta5}. Its domain of validity was studied in the parameter space $(n,\a)$. The values taken here lie within this domain.

In this paper we consider the inverse problem of which cosmology one would expect for a given scalar field profile. This perspective is sometimes used in inflationary or big bang cosmology when one requires a particular profile. For instance, in several inflationary 
models the scalar potential is derived from other known variables (scalar profile $\Phi(t)$
and Hubble evolution $H(t)$) by the Hamilton--Jacobi equations; if one is interested in runaway potentials, the inverse problem selects simple $\Phi(t)$'s and $H(t)$'s from which one can extrapolate the field potential with those characteristics. A third application of the inverse problem is in the context of ekpyrotic/phantom/braneworld dual models, where a duality between physically inequivalent models is established through the scale factor or Hubble parameter. The scalar potential is consequently extracted either from the Hamilton--Jacobi equations or directly from quantities of the dual model (see Ref.~\cite{tria} and references therein).

Here, the main difference with respect to the cases above is that we also extract the Hubble parameter from the diffusion equation for a given solution. The solution is given \emph{ab initio} because the incomplete gamma function is the only profile we know which allows one to easily solve the diffusion equation and (approximately) the nonlinear scalar equation simultaneously via the duplication formulae of Ref.~\cite{cuta5}. The diffusion equation determines univocally a family of Hubble parameters, while the algebraic and analytic properties of the incomplete gamma fix the scalar equation, and hence the potential. The potentials we obtain are rather simple, thus further justifying the method \emph{a posteriori}. The other independent equation of motion (the Friedmann equation) may not be the one of standard general relativity and one will have to reconstruct it. However, this is not the focus of this work.

We discuss the dynamics of two cosmological scalar profiles, one with the cubic potential of the bosonic OSFT tachyon and the other for a toy model with exponential potential. In the former case, the tachyon rolls down the minimum driving a superaccelerating expansion. We assume the scalar field to be the only matter present in the universe and that it sources the cosmological expansion. In general, the expansion can also be driven by gravity itself, like in higher-derivative theories. This sector is beyond our scope because we do not solve all of Einstein's equations.

Although it is not shown whether the solutions thus found have a direct physical application, they are worth recording as they illustrate a methodology for treating nonlocal \emph{nonlinear} equations analytically on a curved (in particular, Friedmann--Robertson--Walker) background. Such is the first motivation for this work. Moreover, they are explicit realizations of the intuitive idea that the cosmological friction heavily affects nonlocal dynamics. This is clear for the solution with cubic potential, which can be confronted with its unbounded Minkowski analog (the string tachyon): the wild oscillations of the Minkowski solution are absent in the cosmological solution because, due to Hubble friction, the field reaches the minimum of its potential in the infinite future.

The present work has been conceived as a follow-up of our line of investigation, and as such it is not meant to be self-contained. We refer to other publications for extensive discussions on issues which may be relevant to nonlocal diffusing theories\cite{cuta2,cuta4,roll,cuta3,cuta5} and nonlocal cosmological models.\cite{cutac,cuta2} Anyway, we wish to (re)make a cautionary remark. This class of cosmological models are considered to be toy simplifications of a more complete scenario. The gravitational sector is not taken into account adequately, since it is basically local; so, one is actually looking at models where nonlocality has not been retained systematically.\cite{cutac,cuta2} This problem is still unresolved in the literature for this type of nonlocal operators. From the point of view of string theory, the issue regards the closed string background, for which there is not as much control and consensus as we have for the open string. From the point of view of nonlocal toy models, the authors of Ref.~\cite{BMS} consider a scenario of pure gravity. Inspired by that, one might try to implement the nonperturbative diffusion method also into the gravitational sector. Unfortunately, there is a technical reason why we cannot do so, namely, that \emph{if} all sectors were dressed with the same nonlocal operators, then the Hubble parameter (friction term in $\B$) would depend on $r$, thus spoiling Eq.~\Eq{q}. A possible way out is to take different diffusion times $r_i$ for each sector of the theory, but we will leave that for the future.\footnote{There is another class of phenomenological nonlocal models, introduced in Ref.~\cite{DW}, which is not inspired by string theory but by quantum field theory loop effects, where nonlocality is encoded in inverse powers of the Laplace--Beltrami operator. By virtue of a field redefinition similar to the one used in $f(R)$ models, one can reduce these models to local ones of scalar-tensor type.\cite{NO,Jhi08} From that point on, the classical analysis is standard. On the other hand, in our case not only do we not yet have an \emph{Ansatz} for the gravity action, but the nonlocality we specialize in does not allow any field redefinition leading to a simple local system (this point was discussed extensively in Refs.~\cite{cuta2,cuta3}).}


\section{Cubic Potential (Bosonic Tachyon)}

The case $(n,\a)=(2,1)$ is rather interesting because it corresponds to the tachyon of bosonic string field theory.\cite{wi86a,KS1,KS2,fuj03} For these values, we can estimate the error of approximation by computing
\be\label{delt}
\Delta_{\rm max}\equiv \mathop{\rm sup}_{z}\Delta(z)\equiv \mathop{\rm sup}_{z}\left|\frac{{\rm LHS}-{\rm RHS}}{{\rm LHS}+{\rm RHS}}\right|\,,
\ee
where LHS and RHS are, respectively, the left- and right-hand side of \Eq{dupli2}. The validity of \Eq{delt} as a robust assessment of the global error has been discussed and proven in Refs.~\cite{cuta2,cuta4,roll,cuta5}. Any solution of the diffusion and scalar equations is easily constrained asymptotically to be such (see these papers for details), so there is no issue about late- or early-time deviations from an exact solution. If they are small, violations of the equations of motion at some intermediate time $t=t_0$ can be taken into account by adding corrective terms to the solution which vanish asymptotically, but in the present case this is unnecessary because the approximate solution $\Phi$ already captures the correct behaviour at all times. 

\begin{figure}
\begin{center}
\includegraphics[width=9cm]{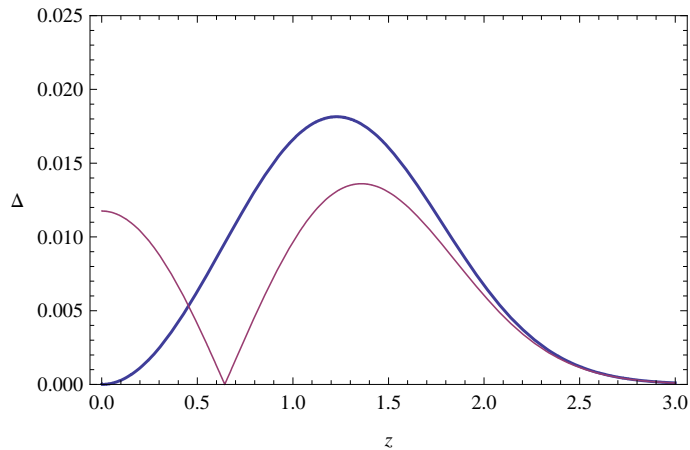}
\end{center}
\caption{\label{fig1} $\Delta$ with the theoretical (thick line) and numerical (thin line) value of $\kappa$. Here $(n,\a)=(2,1)$.}
\end{figure}

This can be seen by looking at the global error associated with $\Phi$, which is small. Plugging in the theoretical value $\kappa=2^{1/4}\approx 1.19$, the supremum norm of $\Delta$ is $\Delta_{\rm max}=\Delta(z\approx 1.23)\approx 1.8\%$. Although this error is already acceptable considering that we have not employed numerical tools, we can improve it by estimating the value of $\k$ which minimizes the global quantity
\be\label{del}
\delta\equiv \mathop{\rm inf}_{\k} \sqrt{\int_0^{\bar{z}} \rmd z \Delta^2(z)}\,,
\ee
where $\bar{z}=O(10)$ is sufficient. We find $\k_{\rm num}\approx 1.18$ and $\delta\approx 1.4\%\approx \Delta_{\rm max}$ (see Fig.~\ref{fig1}).

Since the metric is singular at $t=0$, the solution $\Phi(1,z)=1-\rme^{-z^2}$ is meaningful only at either negative or positive times. The equation of motion is
\be
\ddot\Phi-\frac1t\dot\Phi+\s(\tilde\Phi-\Phi^2)=0\,.
\ee
The static potential $V(\Phi)=\s(\Phi^2/2-\Phi^3/3)$ has a local minimum at $V(0)=0$ and a local maximum at $V(1)=\s/6>0$. Since $-V$ is the potential of the OSFT tachyon (at zero truncation level) on Minkowski, we regard $\Phi$ as a solution of the \emph{Riemannian} problem, i.e., a problem in Euclidean signature on a curved background. The cosmological equations of motion in Euclidean signature are simply those where the usual Wick rotation $t\to \rmi t$ is performed. This interpretation of the equation of motion is necessary only if one identifies the field $\Phi$ with the OSFT tachyon, because the cubic potential is inverted with respect to the usual definition in the theory. If one does not wish to do so, then this problem can be considered to be a toy model in Minkowski signature.

\begin{figure}
\begin{center}
\includegraphics[width=9cm]{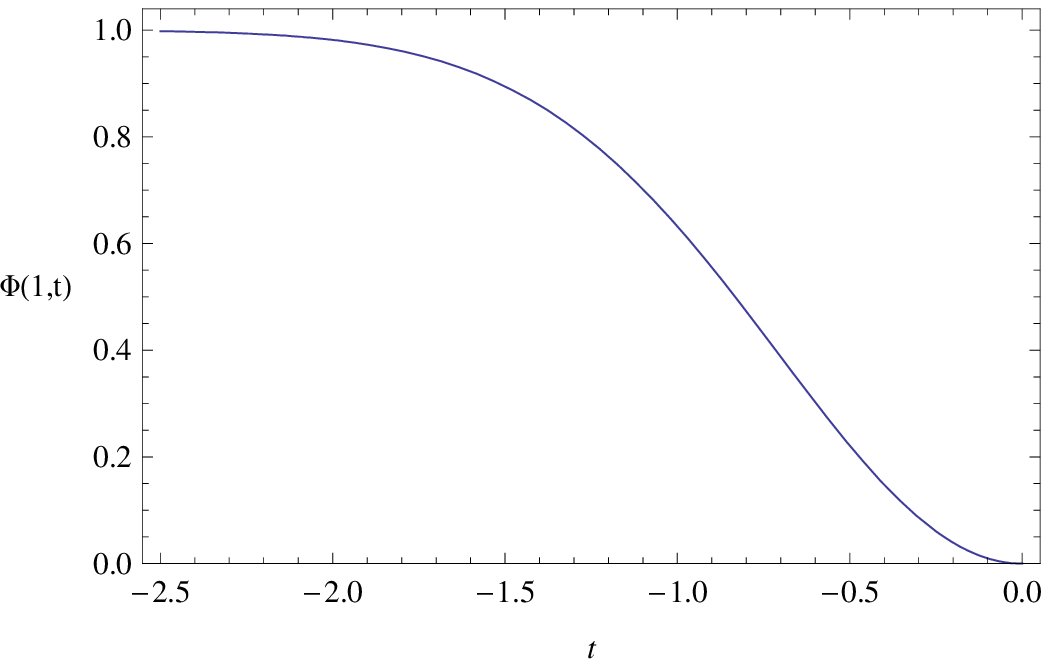}
\includegraphics[width=9cm]{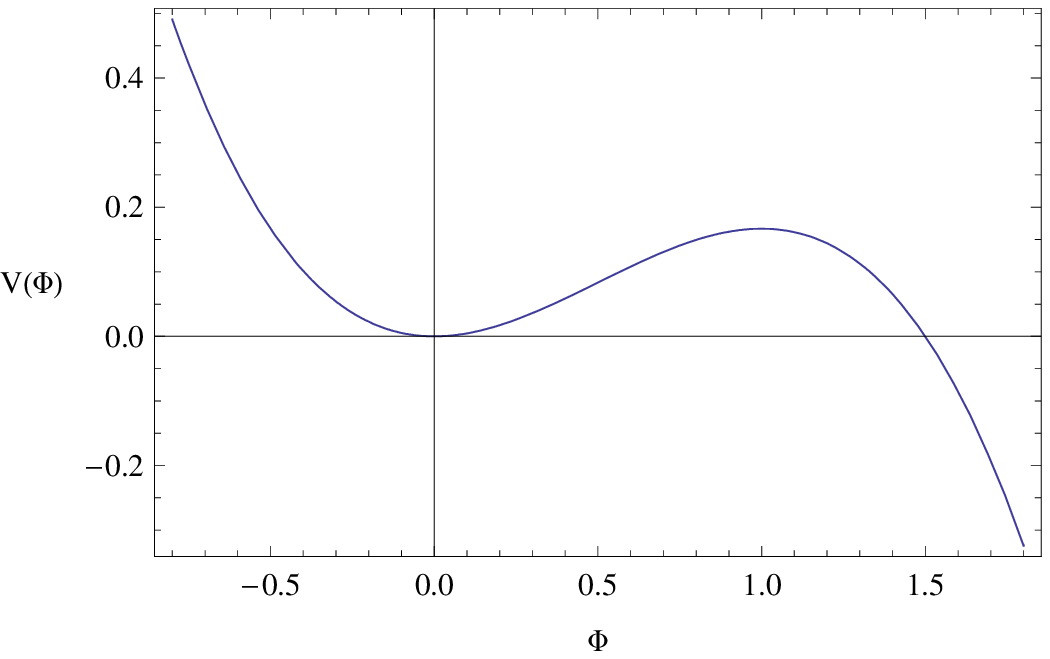}
\end{center}
\caption{\label{fig2} Nonlocal rolling solution (upper panel) and its potential (lower panel) for $(n,\a)=(2,1)$. Here $r=1/4$ and $\s=1$.}
\end{figure}

Let $t<0$. The Hubble parameter is $H\equiv \dot a/a=1/(-3t)$, corresponding to a scale factor $a(t)=(-t)^{-1/3}$. The Riemannian universe expands ($H>0$) in superacceleration ($\dot H>0$). At $t=-\infty$ ($a=0$, big bang) the field sits on the local maximum and starts rolling down the minimum. However, it will never reach it, since the cosmological friction grows with time and the universe will expand indefinitely ($a,H\to+\infty$)\footnote{This is not a big rip singularity\cite{sta00,mci02,CKW,NOT} since it happens at the infinite future where the scale factor is infinite.} (see Fig.~\ref{fig2}). In the absence of a constant term in the potential, the equation of state evolves from a stiff matter regime ($w=p/\rho\sim1$, where $p$ and $\rho$ are, respectively, the pressure and energy density of the scalar field) to a cosmological constant ($w\sim -1$).

In the corresponding local system i.e., the one with $\s$ and $n$ fixed while sending $\kappa$ to zero, if $\Phi$ is solution of the Euclidean (in this case, Riemannian) problem, then $\Psi=1-\Phi$ is solution to the Lorentzian equation of motion. When nonlocal effects are taken into account, this is true only up to a term $\propto(\tilde\Psi-\Psi)$, which vanishes in the local limit. (Nonlocal terms of this form were briefly discussed in Ref.~\cite{cuta2}.) Qualitatively, the dynamics of the nonlocal Lorentzian model with this ``evanescent'' term is the same as the one without it. For $t<0$, $\Psi$ rolls down from the local maximum $V(0)=0$ towards the local minimum at $\Psi=1$, driving superacceleration of an expanding universe.

This behaviour is radically different from the corresponding local picture where the field, like a classical particle subject to friction, oscillates about the Euclidean/Minkowski ($H=0$) minimum with progressively damped oscillations. Although the system is only mildly nonlocal ($s/r\approx 0.29$), the increasing friction in the nonlocal operator eventually forbids the field to ever reach the minimum. Notably, in the absence of friction the nonlocal Lorentzian solution is unbounded, as it undergoes wild oscillations which begin past the minimum.\cite{roll,MZ,CST,KORZ,BMNR}\footnote{The possibility that the Hubble friction be able to damp the wild oscillations was advanced in Refs.~\cite{cutac,FGN} and realized in Refs.~\cite{Jo082,BMNR}. It would be interesting to clarify the relation between the solutions of Refs.~\cite{Jo082,BMNR} and ours.}	

The branch $t>0$ corresponds to a contracting universe with scale factor $a(t)=t^{-1/3}$, evolving towards a big crunch singularity. In this case the scalar field climbs the potential from its local minimum to its local maximum, pulled by antifriction.

Since we have not solved the full cosmological inverse problem, we are not able to assess the physical relevance of this model as far the early history of the universe (inflationary or pre-inflationary) is concerned. However, the first two fast-roll parameters are $\epsilon\equiv-\dot H/H^2=-3$ and $\eta\equiv-\ddot{\tilde\Phi}/(H\dot{\tilde\Phi})=3[1-(\kappa t)^2/(2r)]$, and it would be hard to identify the scalar field as an inflaton generating an almost scale-invariant spectrum. At any rate, this may be regarded as a mathematical example of a bounded OSFT tachyon solution on a curved background.


\section{Exponential Potential}

With little effort, one can obtain a cosmological nonlocal model with an exponential potential. Taking the limit $\a\to 0$ Eq.~\Eq{geom} for $n=1+1/\a$ yields
\be\label{geome}
\B\Gamma(0,\kappa_0^2z^2)=\B\rme^{s\B}\Gamma(0,z^2)\approx\frac{1}{r}\left[\rme^{-\Gamma(0,z^2)}-1\right]\,,
\ee
where $\Gamma(\a,z^2)=\Gamma(\a)-\gamma(\a,z^2)$ is the upper incomplete gamma function, $\kappa_0=\lim_{\a\to0}\kappa=\rme^{\gamma_{\rm EM}/2}\approx 1.33$, and $\gamma_{\rm EM}$ is the Euler--Mascheroni constant. Equation \Eq{geome} is satisfied with the same accuracy as the previous case;\cite{cuta5} therefore we refer again to Fig.~\ref{fig1}.

\begin{figure}
\begin{center}
\includegraphics[width=9cm]{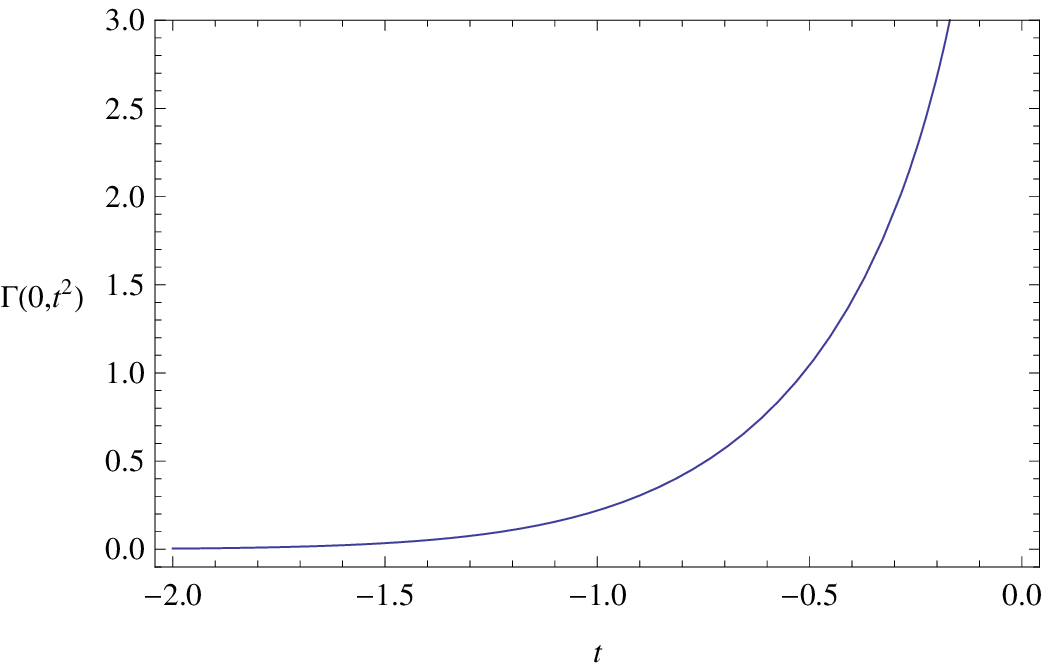}
\includegraphics[width=9cm]{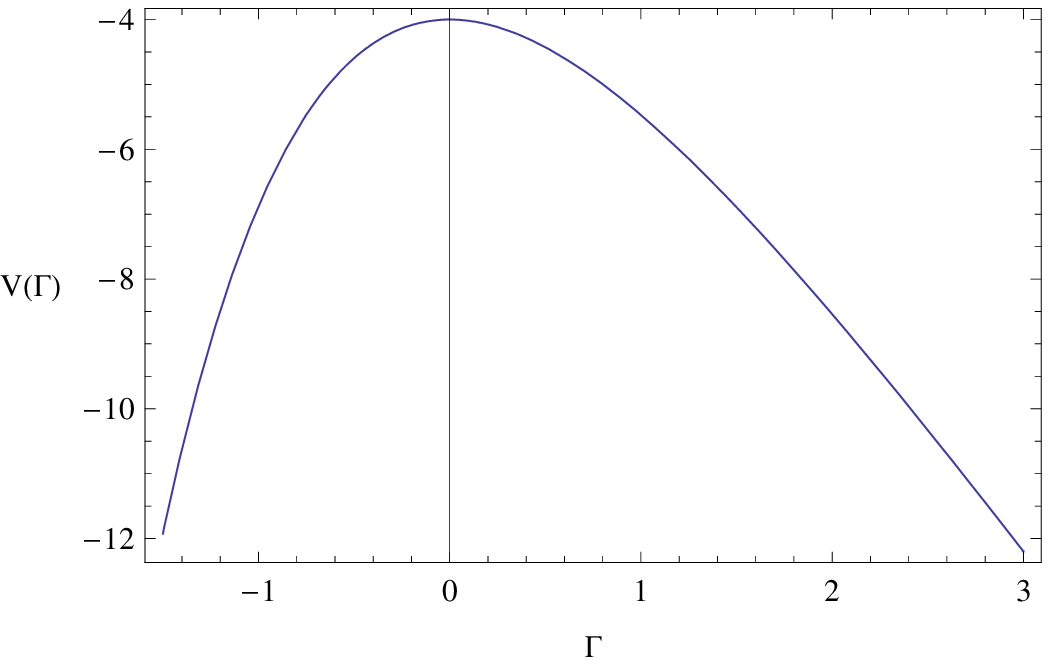}
\end{center}
\caption{\label{fig3} Nonlocal solution (top panel) for an exponential potential (bottom panel). Here $r=1/4$.}
\end{figure}

The potential is $V(\Gamma)=-(\Gamma+\rme^{-\Gamma})/r$. Let $t<0$. The universe contracts with scale factor $a=(-t)^{1/3}$, starting at $\Gamma(t=-\infty)=0$ at the top of the potential $V(0)=-1/r$. Then the field rolls down towards $\Gamma(t=0)=+\infty$, $V(+\infty)=-\infty$ (see Fig.~\ref{fig3}).
The branch $t>0$ represents an expanding universe rolling up to the maximum of the scalar potential i.e., relaxing to the minimum in the Lorentzian case.

To conclude, we have described the dynamics of two models of nonlocal scalar fields on cosmological backgrounds. The scalar with cubic potential drives a phase of superacceleration and may be identified with the bosonic OSFT tachyon. The cosmological friction protects the field from the wild oscillations of its Minkowski counterpart.

\section*{Acknowledgments}

G.~Calcagni is supported by NSF grant PHY0854743, The George A. and Margaret M. Downsbrough Endowment and the Eberly research funds of Penn State. G.~Nardelli is partly supported by INFN of Italy.
 


\begin{thebibliography}{99}

\bibitem{are04} I.Ya. Aref'eva, \textit{AIP Conf. Proc.} \textbf{826} (2006) 301 [arXiv:astro-ph/0410443].
\bibitem{AJ}    I.Ya. Aref'eva and L.V. Joukovskaya, \textit{J.\ High Energy Phys.} \textbf{0510} (2005) 087 [arXiv:hep-th/0504200].
\bibitem{AKV1}  I.Ya.~Aref'eva, A.S.~Koshelev and S.Yu.~Vernov, \textit{Phys.\ Lett.\ B} \textbf{628} (2005) 1 [arXiv:astro-ph/0505605].
\bibitem{cutac} G. Calcagni, \textit{J.\ High Energy Phys.} \textbf{0605} (2006) 012 [arXiv:hep-th/0512259].
\bibitem{AK}    I.Ya. Aref'eva and A.S. Koshelev, \textit{J.\ High Energy Phys.} \textbf{0702} (2007) 041 [arXiv:hep-th/0605085].
\bibitem{AV}    I.Ya. Aref'eva and I.V. Volovich, \textit{Theor.\ Math.\ Phys.} \textbf{155} (2008) 503 [arXiv:hep-th/0612098].
\bibitem{kos07} A.S. Koshelev, \textit{J.\ High Energy Phys.} \textbf{0704} (2007) 029 [arXiv:hep-th/0701103].
\bibitem{AJV}   I.Ya. Aref'eva, L.V. Joukovskaya and S.Yu. Vernov, \textit{J.\ High Energy Phys.} \textbf{0707} (2007) 087 [arXiv:hep-th/0701184].
\bibitem{AV2}   I.Ya. Aref'eva and I.V. Volovich [arXiv:hep-th/0701284].
\bibitem{cuta2} G. Calcagni, M. Montobbio and G. Nardelli, \textit{Phys.\ Rev.\ D} \textbf{76} (2007) 126001 [arXiv:0705.3043].
\bibitem{Jou07} L.V. Joukovskaya, \textit{Phys.\ Rev.\ D} \textbf{76} (2007) 105007 [arXiv:0707.1545].
\bibitem{Jo081} L.~Joukovskaya [arXiv:0803.3484].
\bibitem{ArK}   I.Ya.~Aref'eva and A.S.~Koshelev, \textit{J.\ High Energy Phys.} \textbf{0809} (2008) 068 [arXiv:0804.3570].
\bibitem{Jo082} L.~Joukovskaya, \textit{J.\ High Energy Phys.} \textbf{0902} (2009) 045 [arXiv:0807.2065].
\bibitem{NuM}   N.J.~Nunes and D.J.~Mulryne, \textit{AIP Conf.\ Proc.} \textbf{1115} (2009) 329 [arXiv:0810.5471].
\bibitem{KV}    A.S. Koshelev and S.Yu. Vernov [arXiv:0903.5176].

\bibitem{BBC}   N. Barnaby, T. Biswas and J.M. Cline, \textit{J.\ High Energy Phys.} \textbf{0704} (2007) 056 [arXiv:hep-th/0612230].
\bibitem{lid07} J.E. Lidsey, \textit{Phys.\ Rev.\ D} \textbf{76} (2007) 043511 [arXiv:hep-th/0703007].
\bibitem{BC}    N. Barnaby and J.M. Cline, \textit{J.\ Cosmol.\ Astropart.\ Phys.} \textbf{0707} (2007) 017 [arXiv:0704.3426].
\bibitem{cuta4} G.~Calcagni and G.~Nardelli, \textit{Phys.\ Lett.\ B} \textbf{669} (2008) 102 [arXiv:0802.4395].
\bibitem{BK2}   N.~Barnaby and N.~Kamran, \textit{J.\ High Energy Phys.} \textbf{0812} (2008) 022 [arXiv:0809.4513].

\bibitem{SW}    M.E.~Soussa and R.P.~Woodard, \textit{Class.\ Quant.\ Grav.} \textbf{20} (2003) 2737 [arXiv:astro-ph/0302030].
\bibitem{BMS}   T. Biswas, A. Mazumdar and W. Siegel, \textit{J.\ Cosmol.\ Astropart.\ Phys.} \textbf{0603} (2006) 009 [arXiv:hep-th/0508194].
\bibitem{kho06} J. Khoury, \textit{Phys.\ Rev.\ D} \textbf{76} (2007) 123513 [arXiv:hep-th/0612052].
\bibitem{DW}    S.~Deser and R.P.~Woodard, \textit{Phys.\ Rev.\ Lett.} \textbf{99} (2007) 111301 [arXiv:0706.2151].
\bibitem{NO}    S.~Nojiri and S.D.~Odintsov, \textit{Phys.\ Lett.\ B} \textbf{659} (2008) 821 [arXiv:0708.0924].
\bibitem{Jhi08} S.~Jhingan, S.~Nojiri, S.D.~Odintsov, M.~Sami, I.~Thongkool and 
S.~Zerbini, {\it Phys.\ Lett.\ B} {\bf 663} (2008) 424 [arXiv:0803.2613].
\bibitem{CENO}  S.~Capozziello, E.~Elizalde, S.~Nojiri and S.D.~Odintsov, \textit{Phys.\ Lett.\ B} \textbf{671} (2009) 193 [arXiv:0809.1535].
\bibitem{DeW}   C.~Deffayet and R.P.~Woodard, \textit{J.\ Cosmol.\ Astropart.\ Phys.} \textbf{0908} (2009) 023 [arXiv:0904.0961].

\bibitem{vol03} Ya.I. Volovich, \textit{J. Phys. A} \textbf{36} (2003) 8685 [arXiv:math-ph/0301028].
\bibitem{FGN}   V. Forini, G. Grignani and G. Nardelli, \textit{J.\ High Energy Phys.} \textbf{0503} (2005) 079 [arXiv:hep-th/0502151].
\bibitem{vla05} V.S. Vladimirov [arXiv:math-ph/0507018].
\bibitem{roll}  G. Calcagni and G. Nardelli, \textit{Phys.\ Rev.\ D} \textbf{78} (2008) 126010 [arXiv:0708.0366].
\bibitem{cuta3} G. Calcagni, M. Montobbio and G. Nardelli, \textit{Phys. Lett. B} \textbf{662} (2008) 285 [arXiv:0712.2237].
\bibitem{MuN}   D.J.~Mulryne and N.J.~Nunes, \textit{Phys.\ Rev.\ D} \textbf{78} (2008) 063519 [arXiv:0805.0449].
\bibitem{cuta5} G. Calcagni and G. Nardelli, \textit{Nucl. Phys. B} \textbf{823} (2009) 234 [arXiv:0904.3744].

%
\bibitem{GR}    I.S. Gradshteyn and I.M. Ryzhik, \textit{Tables of Integrals, Series, Products} (Academic Press, San Diego, 2007).
\bibitem{tria}  G. Calcagni, \textit{Phys.\ Rev. D} \textbf{71} (2005) 023511 [arXiv:gr-qc/0410027].

\bibitem{wi86a}  E.~Witten, \textit{Nucl.\ Phys.\ B} \textbf{268} (1986) 253.
\bibitem{KS1}    V.A. Kosteleck\'{y} and S. Samuel, \textit{Phys. Lett. B} \textbf{207} (1988) 169.
\bibitem{KS2}    V.A. Kosteleck\'{y} and S. Samuel, \textit{Nucl. Phys. B} \textbf{336} (1990) 263.
\bibitem{fuj03}  M.~Fujita and H.~Hata, \textit{J.\ High Energy Phys.} \textbf{0305} (2003) 043 [arXiv:hep-th/0304163].

\bibitem{sta00} A.A.~Starobinsky, \textit{Grav.\ Cosmol.} \textbf{6} (2000) 157 [arXiv:astro-ph/9912054].
\bibitem{mci02} B.~McInnes, \textit{J.\ High Energy Phys.} \textbf{0208} (2002) 029 [arXiv:hep-th/0112066].
\bibitem{CKW}   R.R.~Caldwell, M.~Kamionkowski and N.N.~Weinberg, \textit{Phys.\ Rev.\ Lett.} \textbf{91} (2003) 071301 [arXiv:astro-ph/0302506].
\bibitem{NOT}   S.~Nojiri, S.D.~Odintsov and S.~Tsujikawa, {\it Phys.\ Rev.\ D} {\bf 71} (2005) 063004 [arXiv:hep-th/0501025].

\bibitem{MZ}    N.~Moeller and B.~Zwiebach, \textit{J.\ High Energy Phys.} \textbf{0210} (2002) 034 [arXiv:hep-th/0207107].
\bibitem{CST}   E. Coletti, I. Sigalov and W. Taylor, \textit{J.\ High Energy Phys.} \textbf{0508} (2005) 104 [arXiv:hep-th/0505031].
\bibitem{KORZ}  M. Kiermaier, Y. Okawa, L. Rastelli and B. Zwiebach, \textit{J.\ High Energy Phys.} \textbf{0801} (2008) 028 [arXiv:hep-th/0701249].
\bibitem{BMNR}  N.~Barnaby, D.J.~Mulryne, N.J.~Nunes and P.~Robinson, \textit{J.\ High Energy Phys.} \textbf{0903} (2009) 018 [arXiv:0811.0608].
\end{thebibliography}
\end{document}